# An investigation of the relationship between morphology and chemistry of the D-type spherules from the recovery expedition of the CNEOS 2014-01-08 bolide: Implications for origins

Eugenia Hyung[a,c], Juliana Cherston[b,c], Stein B. Jacobsen[a,c], and Abraham (Avi) Loeb[b,c]

[a] Dept. of Earth and Planet. Sci., Harvard Univ., Cambridge, 02138, MA
[b] Dept. of Astronomy, Harvard Univ., Cambridge, 02138, MA
[c] Interstellar Expedition of the Galileo Project, Cambridge, 02138, MA

## Abstract

Cosmic spherules have largely been classified into S-, I-, and G-types according to their compositions, and are identified to have chondritic or achondritic materials as precursors. A recent recovery expedition attempted to sample fragments of the CNEOS 2014-01-08 bolide retrieved roughly 850 magnetic particles, some of which have unknown origins. Among those identified were a new group of highly differentiated materials consisting of close to 160 specimens categorized as "D-type" particles.

We studied the D-type particles with the goal of comparing their various morphological features to their chemical compositional groupings. Four morphological classifications are considered: "scoriaceous," "stubby," "blocky," and "vesicular." The specimens from the "scoriaceous" and "stubby" groups exhibit a spinel/magnetite rim in at least one instance, characteristic of atmospheric entry, and textures indicative of quenching such as dendritic microcrystalline structures, suggesting that a subset of specimens from these groups are candidates for materials of extraterrestrial origin. The particles exhibiting "blocky" and "vesicular" textures are likely to be terrestrial in origin, with no obvious quench features or signs of ablation. The D-type particles identified and characterized in this study have a spectrum of terrestrial and probable extraterrestrial origins.

## Introduction

Cosmic spherules refer to micrometeorites that are primitive in composition and have undergone significant degrees of melting during exposure to high temperatures during atmospheric entry. Cosmic spherules are largely classified into S-type, I-type, and G-types based on their compositions, with S-types being "stony," or silicate-based, I-types being iron-rich, and G-types being of a composition between the two (Genge et al., 2008). In addition to chondritic precursors, some micrometeorites and cosmic spherules have been identified to be differentiated, originating from the achondrite HED class of meteorites from Vesta (Brase et al., 2021; Cordier et al., 2012, 2011; Soens et al., 2022; Taylor et al., 2007).

Highly differentiated compositions have not been documented among cosmic spherules. Fractionated REE patterns have been used as a criterion for discerning highly differentiated materials to be of terrestrial origin (Folco and Cordier, 2015). Terrestrial spherules have been identified to have several different origins, these largely being volcanic (Amonkar et al., 2021;

Goresy, 1968; Iyer et al., 1997a, 1997b; Porritt et al., 2012), lightning-induced (Genareau et al., 2015; Genareau et al., 2019a; Genareau et al., 2019b), and impact-related (Glass and Simonson, 2012; Simonson and Glass, 2004). Volcanic spherules have been ascertained as such due to the presence or enrichment in magmaphile elements (Iyer et al., 1997a) and enrichment of REEs (Folco and Cordier, 2015). Compositional similarities to the upper continental crust, glassy textures (Glass and Simonson, 2012), morphological indications of high pressure impact such as concentric zoning and inward radiating crystallization textures (Glass and Simonson, 2012), features such as lechatelierites (Glass and Simonson, 2012; Simonson and Glass, 2004) and schlieren (Folco et al., 2010; Simonson and Glass, 2004; Yan et al., 2022) are characteristic of microtektites, or impact spherules. Compositional enrichment in Si compared to Fe and Al, and cracked surfaces are characteristic of lightning-induced particles (Genareau et al., 2015). Spherules that are volcanic, and are of volcanic lightning-induced origins are understudied (Wozniakiewicz et al., 2024).

The investigation of morphologies and textures of micrometeorites have been essential to their interpretation, as such features often preserve clues to their thermal history and origins (Genge et al., 2008; Van Ginneken et al., 2017). Micrometeorites are classified into three groups according to varying degrees of thermal alteration: melted micrometeorites, partially melted micrometeorites, and unmelted micrometeorites (Folco and Cordier, 2015). Melted micrometeorites, or cosmic spherules, are most often spherical from the surface tension of the melts (M. Van Van Ginneken et al., 2024) and are further categorized according to their textures that reflect different degrees of heating, although their parent materials can also have an effect (Van Ginneken et al., 2017). S-type spherules are further grouped into transparent and glassy spherules (V-type), cryptocrystalline spherules (CC), barred olivine spherules (BO), porphyritic spherules (PO) the order of decreasing peak temperatures during atmospheric entry (Folco and Cordier, 2015; Genge et al., 2008; Taylor et al., 2000). Partially melted micrometeorites, known as scoriaceous micrometeorites, are exposed to lower temperatures than melted micrometeorites. They are irregular (subrounded) in shape and may preserve relict grains (Folco and Cordier, 2015). Magnetite/spinel or igneous rims are characteristic of partially melted and unmelted micrometeorites, and have been used as a diagnostic feature of extraterrestrial origins (Folco and Cordier, 2015; Genge et al., 2008; Toppani et al., 2001; Toppani and Libourel, 2003) and observed in impact spherules. Dendritic microcrystalline magnetites (Genge et al., 2008; Iwahashi, 1991; Pandey et al., 2023; Taylor et al., 2000) are evidence of quench conditions and are commonly observed in micrometeorites and cosmic spherules; however, these features are not unique to extraterrestrial materials and can be observed in the mesostasis of volcanic materials have undergone quench conditions. Unmelted micrometeorites are a class of micrometeorites that have undergone the lowest degree of thermal alteration, are irregular, and similar to partially melted micrometeorites, exhibit magnetite/spinel or igneous rims, and dendritic microcrystalline magnetites.

A recent expedition that was aimed at collecting possible fragments of the CNEOS 2014-01-08 bolide, collecting close to 850 particles (Loeb et al., 2024), using a sled covered with neodymium magnets. Based on their bulk composition, particles were determined to be cosmic in origin with chondritic precursors and were classified into the archetypal S-type, I-type (high-Ni, low-Ni), and G-type spherules. Of the samples collected from the expedition, roughly ~20% of these were judged to deviate from typical primitive compositions resembling chondrites and newly classified as "D-type" particles. The high degree of differentiation was apparent from their low Mg contents and their highly fractionated REE patterns and elevated elemental abundances. While differentiated particles have typically referred to micrometeorites of the achondrite class of bodies, the D-type particles were demonstrated to be distinct from achondrites in their degree of differentiation (**Figure 1**), and therefore labeled as such to distinguish these materials from other micrometeorites of cosmic origin. Among the sub-class of D-type particles, "BeLaU" spherules, named for their elevated abundances in highly incompatible elements such as Be, La, and U, were classified to be of unknown origin (Loeb et al., 2024). One BeLaU spherule, whose polished cross-section was analyzed, exhibited a magnetite rim and quench features, suggestive of an extraterrestrial origin.

Whereas the study of Loeb et al., (2024) have focused on the BeLaU particles among the D-type particles, here we study and classify an additional subset of the D-type particles from this collection. We aim to determine whether morphology correlates with bulk composition, and whether the combination of composition and texture may be diagnostic of their origins. Backscattered electron images were obtained on polished cross-sections to observe their inner features, and elemental analyses were performed to determine the compositions of the various features. The study aims to use morphological features and compositions to further classify the D-type particles in association with their compositions, and to categorize samples of interest for future studies aimed at exploring their possible extraterrestrial origins.

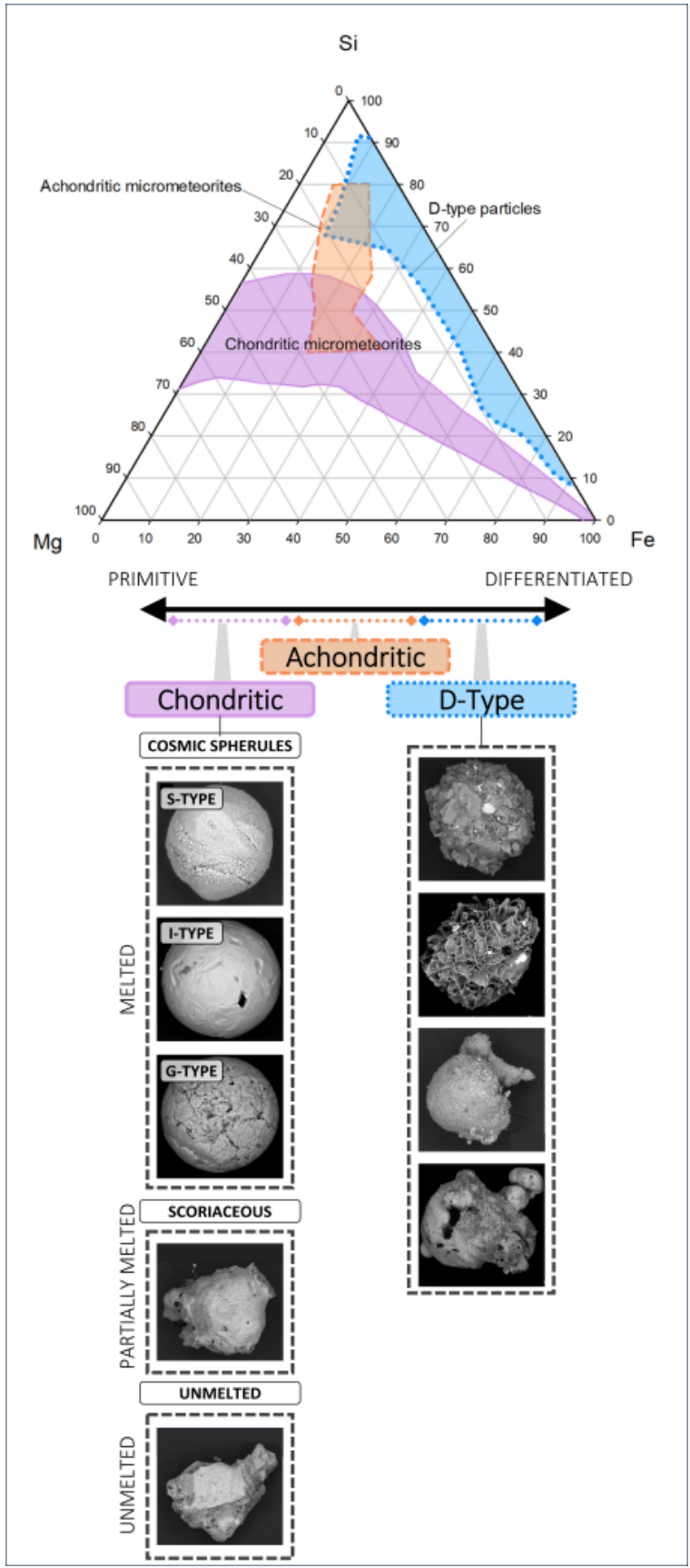

**Figure 1.** Top: Ternary diagram of the Mg-Si-Fe contents of the cosmic spherules of chondritic (solid line; after Brase et al., 2021; Soens et al., 2022) and achondritic origins (dashed line, after Soens et al., 2022), plotted alongside D-type particles (dotted line; Loeb et al., 2024). Bottom: Diagram depicting the classification of different cosmic spherules according to composition and morphology in relation to D-type particles.

## Methods

Whole-sample bulk major element data was retrieved from micro-XRF analysis as discussed in Loeb et al. (2024). In the study, D-type spherules were further categorized into four subgroups according to their Sr and Si contents made through these analyses. These are the "low-Sr, low-Si," "low-Sr, high-Si," "high-Sr, low-Sr," and "high-Sr, high-Si" groups. The cut-off between the low-Sr and high-Sr groups is at 450 ppm. While the cut-off for the low-Si and high-Si groups was determined to be $100*Si/(Si+Mg+Fe) = 70$ in molar fractions for the low-Sr group, cut-off Si content for the high-Sr D-type particles is different, at 60 (Loeb et al., (2024).

Among the ~850 samples from the expedition, ~160 were classified as D-type spherules. Thirty-seven of the ~160 were imaged for backscattered electron images and EDS analysis. The BSE images were used to categorize their particles according to their morphology. To further investigate whether the textures on the surface of the particles are primary features or alteration products, polished cross-sections were made of four specimens for analysis, one from each morphological group.

A JEOL JXA-8230 electron microprobe was used to obtain backscattered electron (BSE) images, elemental X-ray maps, and chemical analyses by energy (EDS) and wavelength (WDS) dispersive spectroscopy. A focused beam of 1 μm in diameter was used. BSE images were obtained on samples mounted on top of an acrylic slide with transparent double-sided tape. The samples were carbon-coated prior to analysis. The chemical compositions of the surfaces of the particles were measured by EDS using factory calibration curves, while polished cross-sections were analyzed by WDS using natural minerals and synthetic glasses as calibration standards (Petaev and Jacobsen, 2009). EDS analyses and imaging were performed at an accelerating voltage of 20 kV and beam currents of 5–10 nA and ~0.1 nA at low- (< 1000×) and high-resolution (> 2000×), respectively. An accelerating voltage of 15 kV was used for WDS analysis, along with a beam current of 20 nA. Counting times of 30 sec and 15 sec on the peak and background, respectively, were used.

**Results**

*Morphology*

In addition to the four compositional subclassifications proposed by Loeb et al., (2024) which further divide the D-type particles according to their Si and Sr-contents, the D-type spherules/particles were further subclassified into four categories based on their morphological characteristics. The four major subgroups are "blocky," "scoriaceous," "vesicular," and a category for the particles that exhibit textures resembling microcrystalline features on their surface, which we refer to as "stubby" throughout this study (**Figure 2**). We note that these are descriptors which have been or may be used to characterize both terrestrial samples and cosmic spherules. These classifications were determined according to their observed surficial morphologies during BSE image analysis. Categorization of the particles without BSE images is difficult, as some textures and surficial features are subtle under a microscope and obscured by lighting. Therefore, only a fraction of the D-type spherules was able to be categorized into four major groups with certainty according to their most prominent features observed in the BSE images. The D-type particles investigated in this study exhibited shades of grey or black under a microscrope. A small subset of the D-type spherules is identified to be spherical or subspherical, with rugged surface textures that resemble those of other scoriaceous D-type particles. A separate group, "shards/fragments," is reserved for particles that were fragmented to the extent that their original morphology is unclear. The fragmented surfaces of these shards are observed to exhibit dendritic microcrystalline structures (17MAG-35 and IS8M2-9; SI).

*Analysis of specimens from each morphological subtype*

Scoriaceous

The particles characterized to be "scoriaceous" are labeled after their irregular morphologies. While the scoriaceous particles from the class of D-type spherules are reminiscent of partially melted micrometeorites in terms of similarities demonstrated in their morphology, such a texture is not unique to micrometeorites and are commonly observed in terrestrial lavas that exhibit porous textures due to volatile loss during decompression and cooling. Scoriaceous micrometeorites are also similarly characterized by rough, uneven surfaces and asymmetrical, irregular (subrounded) shapes, and vesicles. A small fraction of the scoriaceous particles displayed in this study (**Figure 2c, d**; SI for zoomed images) exhibited dendritic patterns, indicative of quench conditions. Some shards from this study, whose original morphologies are unrecognizable, exhibit dendritic patterns on their fragmented surfaces, similarly to **Figure 2c** (see also the SI for zoomed images).

The polished cross section of one scoriaceous particle (17NMAG-28, high-Sr, low-Si) reveals a large concavity in addition to Fe-Al spinels arranged in a dendritic microcrystalline pattern and larger Fe-Al-oxide relict grains (**Figure 3b, c**; Loeb et al. 2024), in a glassy aluminosilicate mesostasis. No other distinct phases are apparent. A thin, partial Fe-Al spinel rim is observed (**Figure 3b**). The mesostasis consists of Al, Si, Na, Mg, K, Ca, and Fe. Three different classes of D-type particles were observed to exhibit scoriaceous textures, which are the high-Sr, low-Si; low-Sr, low-Si; and high-Sr, high-Si groups.

Stubby

The stubby particles were grouped due to small bumps observed on their surfaces. Upon close investigation, many of the BSE surface images of the stubby particles exhibit microcrystalline structures, with textures that are suggested to be dendritic (**Figure 2g, h, and l** corresponding to samples IS19M-18, 19MAGx-4, IS19M-16 respectively; SI). In particular, 19MAGx-4 has been identified as a BeLaU spherule in Loeb et al., (2024). A cross-section BSE image of sample IS19M-12 (low-Sr, low-Si) indicates that these features extend into the interior of the particle (**Figure 3e**). Relict grains consisting of Fe-Al-spinel are observed (**Figure 3e and f**). The aggregate of the relict grains appears to resemble a spherical core or bead in the cross-section BSE image (**Figure 3e**). The darker mesostasis is composed of aluminosilicate glass and consists of dendritic microcrystalline textures composed of Fe-Al spinel. Similarly to particle 17NMAG-28, no other distinct phases are apparent. Lighter, dendritic textures consisting of Fe-Al spinel are also observed near the surface of the specimen. Small vesicles are occasionally observed throughout the cross-section.

Blocky

The particles categorized as having "blocky" morphologies are, as suggested by the name, characterized to be coarse grained, consisting of blocky aggregates with relatively sharp edges. Titanomagnetite crystals were observed on the surface of some of the particles (light-colored grains in **Figure 2n, o, p, q, and r**). The cross-section analysis of sample IS12M3-12 (low-Sr, high-Si) revealed a large concavity in the middle, with identified minerals such as plagioclase, olivine, pigeonite, and titanomagnetite (**Figure 3i**) in a glassy mesostasis. No quench features or indications of ablation were observed. *The blocky spherules analyzed in this study are, to a large extent, observed to plot within the domain which outlines the most common terrestrial samples in a Mg-Fe-Si ternary diagram.* Most of the blocky D-type particles investigated in this study were determined to be from the high-Sr, high-Si group, with the two samples identified to be from the low-Sr, high-Si group.

Vesicular

The "vesicular" particles have been grouped according to the abundance of vesicles observed on the surface of the samples. While vesicles are observed throughout various other morphologies of D-type particles and in micrometeorites in general, in this study, this term is used to subcategorize a group of D-type particles whose features are particularly common throughout almost the entirety of the sample, starting from the surface (**Figure 2s–x**). A BSE image of the polished cross-section of particle IS18M-3 (low-Sr, low-Si; **Figure 3k–l**) reveals the bulk of the material to consist of glass, consisting of Na, Mg, Al, Si, K, Ca, Ti, and Fe, with Si being the most abundant (**Table S1**). The protruded crystal from the same cross-section (**Figure 3l**) is characterized as a titanomagnetite. Unlike the scoriaceous and stubby particles, quench features such dendritic magnetite and spinels are absent from this group. Many of the vesicular particles consist of the low-Sr, high-Si, and low-Sr, low-Si subclassifications of D-type particles (**Figure 4d**).

*Composition*

For comparison, the D-type particles are plotted in a ternary diagram (**Figure 4**), where the sum of Mg, Si, and Fe contents of the materials have been normalized to 100 in molar proportions. As this kind of plot makes limited use of the space, a second ternary diagram is composed of Sr, Si, and Fe, is used (**Figure 5**). Similarly to Loeb et al. (2024), the Sr, Si, and Fe contents are normalized with respect to CI chondrites (Anders and Grevesse, 1989) in ppm, and each normalized element is further normalized to the sum of their values in % such that:

$$Sr = \frac{Sr_{CI}}{Sr_{CI} + Si_{CI} + Fe_{CI}} \times 100$$

The normalized Si and Fe contents are additionally multiplied by 100 to be able to make efficient use of the space comprising the ternary diagram:

$$Fe = \frac{100 Fe_{CI}}{Sr_{CI} + Si_{CI} + Fe_{CI}} \times 100$$
$$Si = \frac{100 Si_{CI}}{Sr_{CI} + Si_{CI} + Fe_{CI}} \times 100$$

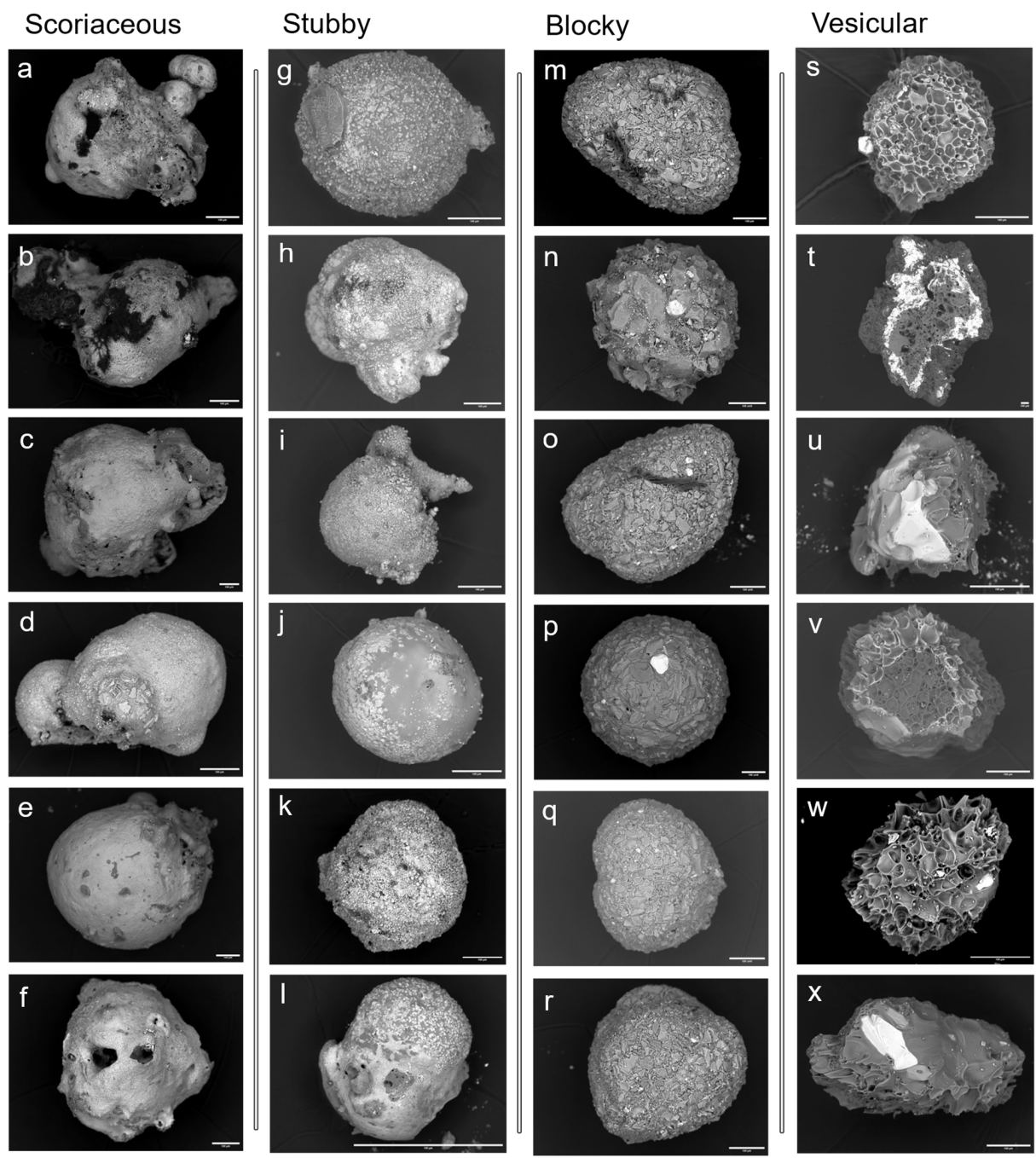

**Figure 2.** Nonspherical D-type particles categorized into four major groups according to their morphologies from top to bottom, left to right: (a–f) scoriaceous (g–h) stubby (m–r) blocky and (s–x) vesicular. Scale bars indicate 100 µm.

**Table 1.** Number of particles tallied according to their observed morphologies and chemical compositions.

|                   | scoriaceous | stubby | blocky | vesicular | fragments |
|-------------------|-------------|--------|--------|-----------|-----------|
| low-Sr, low-Si    | 1           | 3      | –      | 8         | –         |
| low-Sr, high-Si   | –           | –      | 2      | 2         | –         |
| high-Sr, low-Si   | 6           | 3      | –      | –         | 3         |
| high-Sr, high-Si  | –           | 1      | 10     | 1         | –         |

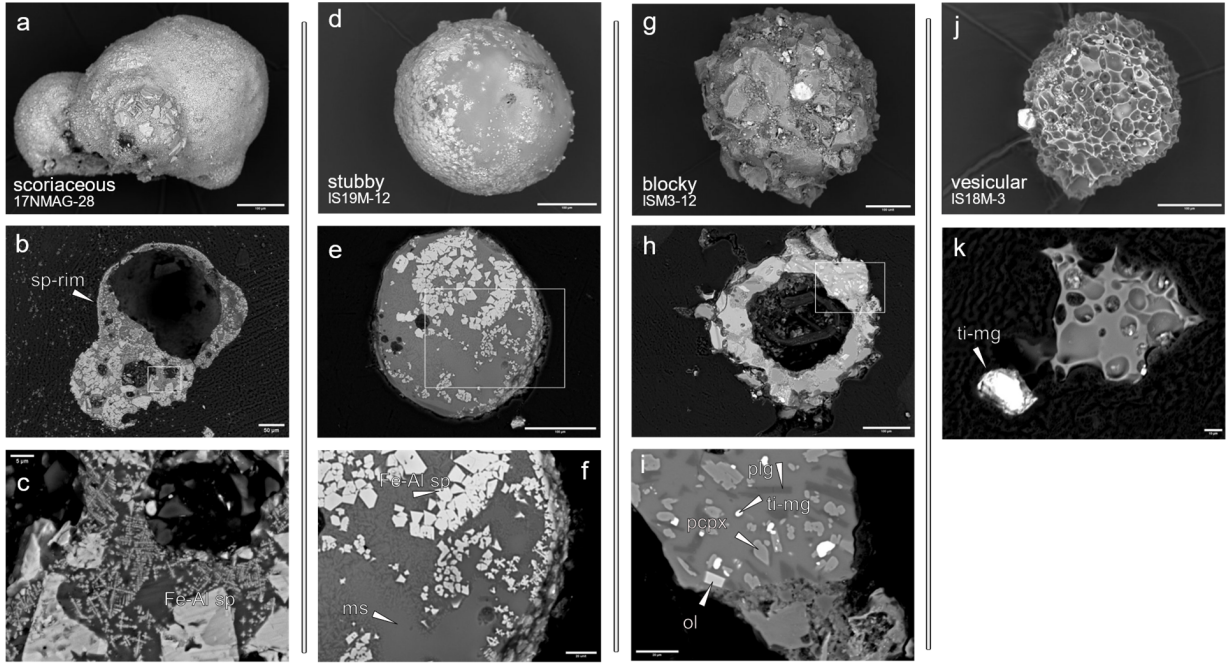

**Figure 3.** Cross section BSE images of particle representing each of the four groups of D-type particles classified according to their morphologies from top to bottom, left to right: (a–c) scoriaceous (17NMAG-28) (d–f) stubby (IS19M-12) (g–i) blocky (IS12M3-12) and (j–k) vesicular (IS18M-3). Panels c), f), and i) represent zoomed-in images of the insets in b), e), and h), respectively. b) sp-rim: Fe-Al spinel rim. A close-up view of the **rims and EDS data** are available in Loeb et al., (2024). c) The relict grains composed of Fe-Al spinel exhibit zoning, where the different dark and bright patches reflect varying degrees of Fe and Al. e) The lower, right side of the particle that is exposed from the epoxy, indicates that the microcrystalline textures on the surface extend into the inner parts of the particle. f) Fe-Al spinel relict grains are present in an alumnosilicate mesostasis (ms). i) plg: plagioclase; ti-mg: titanomagnetite; pcpx: pigeonite; ol: olivine. k) the arrow points to the titanomagnetite (ti-mg) crystal also shown in panel j). Scale bars indicate 100 μm for a), d), e), g), j) and h); 50 μm for b); 20 μm for f) and i); 10 μm for k); and 5 μm for c). Spot analyses data are available in the supplement.

# Discussion

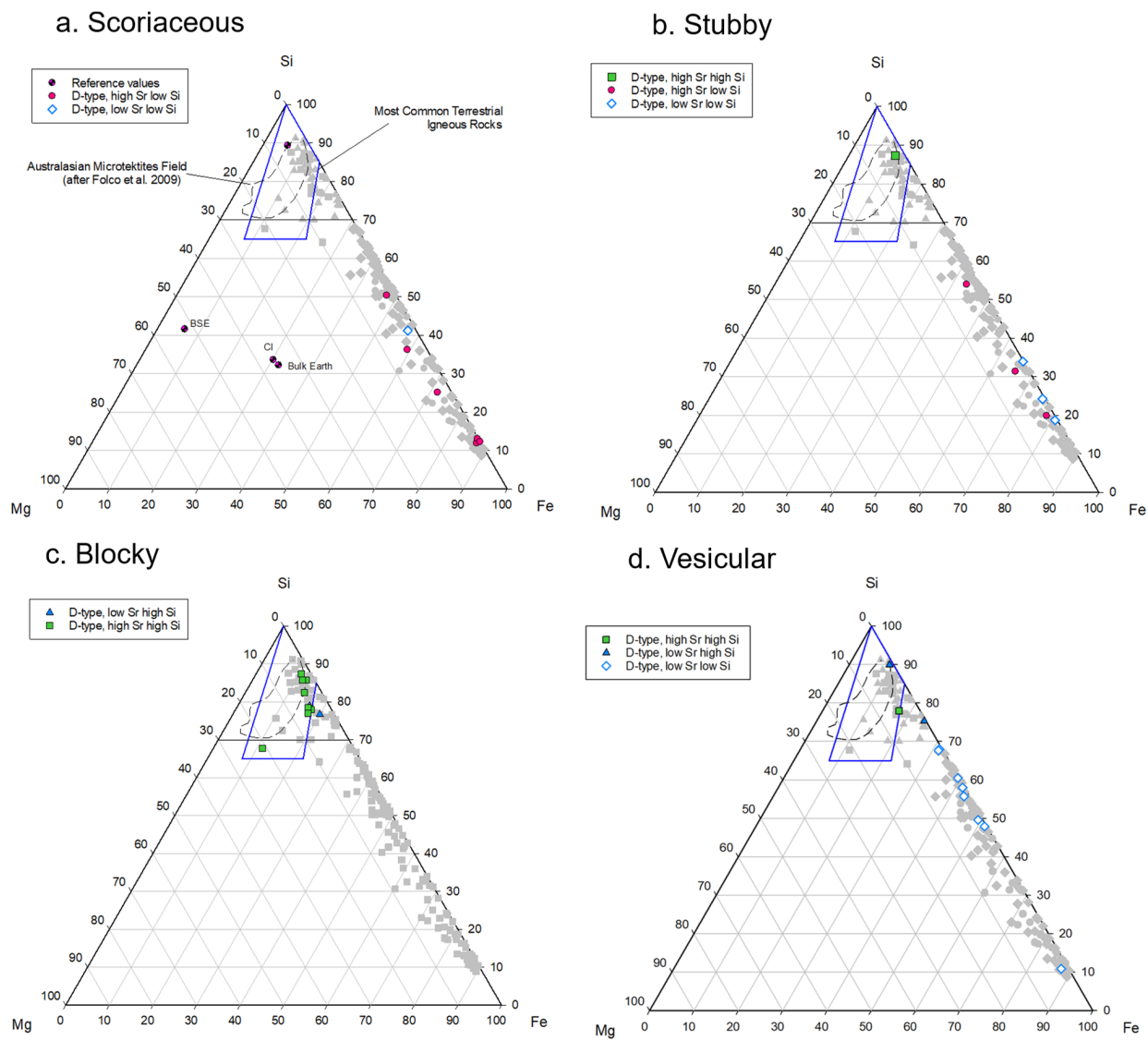

**Figure 4.** Mg-Si-Fe ternary diagrams of the different classes of spherules, categorized according to their morphologies, where the sum of the Mg, Fe and Si contents of the particles is normalized to 100 in mole fractions. The line delineates the distinction between high-Si and low-Si D-type spherules (for low-Sr spherules) following Loeb et al. (2024). The reference materials plotted in a) are the Bulk Earth and BSE (bulk silicate Earth; McDonough and Sun, 1995), UCC (upper continental crust; Rudnick and Gao, 2003), and CI (CI chondrites; Anders and Grevesse, 1989). The boundaries of the Australasian microtektite field are delineated in a dashed line (Brase et al., 2021; Folco et al., 2009; Glass et al., 2004; Van Ginneken et al., 2018). The blue solid line delineates the most common terrestrial rocks (after Loeb et al., 2024). The greyed shapes plot all the D-type particles identified from the expedition, while the shapes with outlines and color are specimens that have been characterized for their morphologies with BSE images using an EPMA. Bulk compositional data are from Loeb et al., (2024).

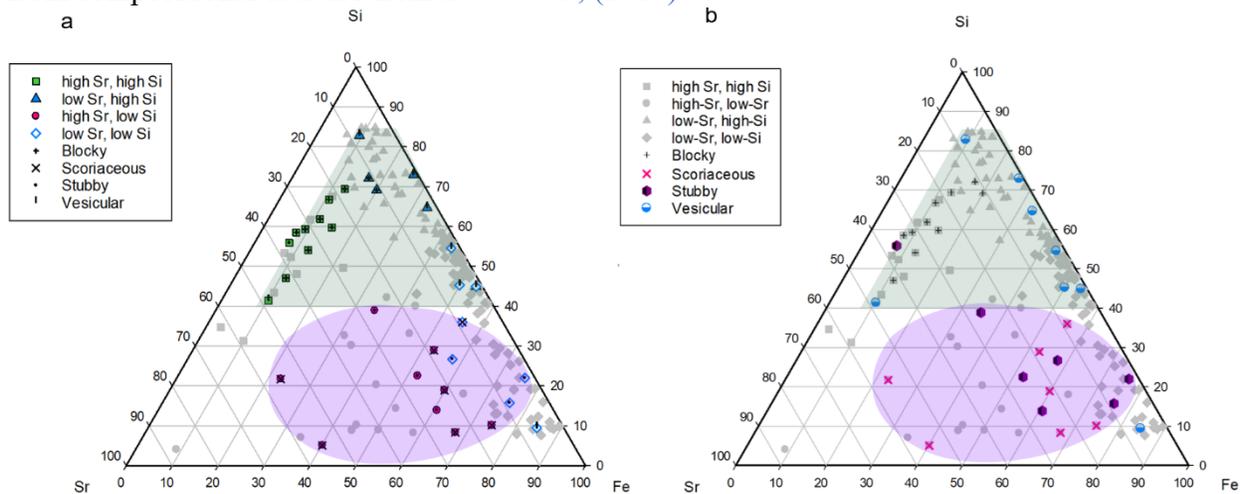

**Figure 5**. Different classes of spherules plotted with respect to their Sr, Si, and Fe contents. In general, the greyed shapes represent all the particles identified to be D-type particles, while the samples whose morphologies have been characterized through BSE images are outlined or represented in colored symbols. The different shaded areas highlight the domain that the blocky and vesicular particles take up (green) and the domain that consists of scoriaceous and stubby particles (purple). Panel a) highlights the subclassifications of D-type compositions in color, while panel b) highlights the different morphologies.

*Terrestrial spherules*

The particles exhibiting blocky textures are plotted in a Mg-Si-Fe ternary diagram (**Figure 4**) for comparison and analysis. The blocky particles are suggested to be most like the most common terrestrial rocks in composition in terms of their Mg-Si-Fe contents among the different morphological groups. In turn, most of the high-Sr, high-Si D-type particles investigated in this study are characterized to have a blocky texture.

Blocky textures have previously been observed in terrestrial spherules of hydrothermal and volcanic origins. The textures exhibited and described in studies such as Amonkar et al., (2021) are less pronounced, as the features may have been etched and eroded in the deep sea environment from which the specimens were collected, reflecting the greater degree of weathering involved compared to the particles in this study. Blocky textures are not entirely limited to D-type particles. Micrometeorites in other collections have been described to exhibit this texture (Blanchard et al., 1980; Noguchi et al., 2017), as many cosmic spherules can be porphyritic due to crystals that form during atmospheric entry. Thus, while blocky morphologies seem to be common in materials of terrestrial origin, this surficial texture alone is not a suitable heuristic for discerning materials of terrestrial versus extraterrestrial origins.

Microtektites are another point of comparison. Owing to the location where the D-type particles were collected, the group of microtektites that are most comparable to these samples are from the Australasian strewn field. The Australasian strewn field is the largest of known strewn fields, covering at least 10% of the Earth's surface (Folco et al., 2010) extending to Australia, Southeast Asia, and parts of Antarctica. Contamination from the impactor has been discerned by elevated Cr, Ni and Co abundances, Cr/Ni and Co/Ni ratios and Cr/Ir ratios (Folco et al., 2023, 2018). Recent calculations have yielded impactor contributions of up to ~6% to microtektites (Folco et al., 2023) which suggest that the contribution of the impactor material to microtektites are small. Thus, in a Mg-Si-Fe ternary diagram composed of major elements, the domain formed by Australasian microtektites primarily reflect the composition of target rock (the upper continental crust) rather than the impactor (Glass and Simonson, 2012), which overlaps with the domain constituted by the most common terrestrial rocks (**Figure 4a**).

While the impact site for the Australasian strewn field is currently unknown, its putative location is suggested to be around Indochina (Glass and Pizzuto, 1994; Lee 2006; Wei, 2000; Prasad et al., 2007). Microtektites collected from a distance that exceed 10 crater diameters consist of glassy, spherical morphologies rather than fine- or course-grained textures (Glass and Simonson, 2012) where mineral inclusions (other than lechatelierites) tend to become rarer. While the blocky particles, as well as the high-Sr, high-Si subclass of D-type particles, are suggested to be similar in composition to the Australasian microtektites, glassy textures would be expected considering the distance of the expedition site from the estimated impact site for the Australasian strewn field. Based on morphology, it is deduced that these materials are unlikely to be microtektites. These particles are similarly difficult to classify as microkrystites (impact spherules that contain primary microlites, which often consist of clinopyroxene; Glass and Burns, 1988), due to the generally nonspherical and rugged surface, which lacks a glassy texture.

The blocky D-type particles of this study and Loeb et al., (2024) lack dendritic patterns and magnetite rims, indications of quench conditions. The presence of minerals and the blocky aggregates suggests conditions for mineral growth during cooling of a melt. The bulk compositions of these particles plotted in Mg-Si-Fe ternary diagrams in **Figure 4c** indicate similarities to the domain formed by the most common terrestrial magmas, rather than cosmic spherules as indicated in **Figure 1**. The presence of titanomagnetite grains in a number of samples (**Figure 2n, o, p, q, and r**) is suggestive of formation in a terrestrial environment. These features suggest that sample IS12M3-12 (**Figure 3g, h, and i**) may have originated from a typical basaltic magma. The D-type particles characterized as "blocky" in this study (**Figure 2** and **Figure 4**), largely encompassing the high-Sr, high-Si subclass of D-type particles, are likely to be of terrestrial and magmatic origin.

*Volcanic origins*

In contrast to the blocky spherules, the glassy matrix of the vesicular particles suggest that these were cooled relatively rapidly compared to the previous group of particles. In addition, the preservation of subtle vesicular textures, the similarity of the vesicular texture to basalts, and lack of signs of atmospheric heating and quenching suggest that the vesicular D-type particles of this study are likely terrestrial and magmatic in origin. A larger specimen (sample IS17M-19; low-Sr, high-Si; **Figure 2t**) which was characterized to be a high-Sr, high-Si particle, was identified to be a basalt from their morphology and textures. The presence of minerals such as plagioclase, clinopyroxene, titanomagnetite, their similarities in bulk composition to the most common terrestrial rocks in a Mg-Si-Fe ternary diagram, textures resembling basalts, and the lack of quench textures due to frictional cooling support this origin.

Many of the vesicular particles exhibit little to almost no Mg and are high in Fe content, plotting outside of the domain created by the most common terrestrial rocks in a Mg-Si-Fe ternary diagram. While Fe-rich lavas are rare on the Earth and are generally indicative of highly evolved melts, the extent of iron enrichment through partial melting and fractionation cannot be the primary driving factor of this Fe-enrichment as the partitioning behaviors of Fe and Mg only differ in their degree of compatibility while behaving similarly compared to incompatible elements. The Fe-rich compositions of many of the vesicular particles in this collection resemble the iron-rich terrestrial spherules collected from the Central Indian Ocean (Amonkar et al., 2021; Iyer et al., 1997a, 1997b), where some spherules were characterized to consist of up to 76% FeO (Iyer et al., 1997a, 1997b).

Despite the similarly high Fe-content of the D-type particles in this study to those observed in Amonkar et al., (2021) and Iyer et al., (1997a,b), some differences are apparent. The particles in this study lack a well-defined spherical morphology compared to those present in Amonkar et al., (2021), or the smoothness observed among volcanic spherules studied in Iyer et al., (1997a). CIPW norm calculations using spot analyses (SI Table S1, S2) of sample IS18M-3 (**Figure 2s, 3j, k**) and sample IS22M1-13 (**Figure 2w**, SI Table S3, S4) constitute minerals such as quartz, plagioclase, orthoclase, clinopyroxene, orthopyroxene, and ilmenite, typical mineral assemblages of a terrestrial basalt. As the vesicular surface textures resemble those of terrestrial basalts, it is possible that these are fragments of larger basaltic samples. The samples may reflect extreme heterogeneity at this scale, deviating from the bulk compositions of the most common terrestrial rocks in terms of their Mg-Si-Fe contents. This also suggests that the Fe-rich particles in this study may be reflective of sampling bias due to the use of a neodymium magnet sled that was used for sample retrieval.

While these scoriaceous particles also resemble achneliths, or spherical pyroclasts from volcanic eruptions based on morphology alone, their compositions do not seem to resemble typical achneliths. The compositions of achneliths resemble those of terrestrial lavas and are basaltic or basaltic andesite in composition, where chemical variability tends to lean towards an enrichment in $SiO_2$ (up to 98 wt%; Moune et al., 2007). Silica-undersaturated, nephelinitic achneliths exhibit lower amounts of silica compared to basaltic compositions (Carracedo-Sánchez et al., 2016), which still exceed 40 wt%, and demonstrate a moderate amount of MgO present in the sample (~10 wt%). In contrast, the scoriaceous D-type particles of this study demonstrate a relative enrichment in Fe compared to Mg and Si, and are demonstrated to be distinct in comparison to silica-undersaturated, basaltic, and andesitic compositions included in the domain formed by the most common terrestrial magmas outlined in **Figure 4a**.

*Exploring possible extraterrestrial origins for a subset of D-type spherules*

While the stubby particles comprise a separate subcategory in this study based on their morphological features, their bulk compositions exhibit some similarities to the scoriaceous D-type particles (**Figure 4a, b**). Eight of the ten particles characterized as either "scoriaceous" or "stubby" consist of D-type particles characterized to be of the high-Sr, low-Si and low-Sr, low-Si subclasses. It is worth noting that two D-type particles that were classified to be fragmented/shards also exhibited dendritic, microcrystalline structures on some fragmented surfaces (samples 17MAG-35 and IS8M2-9; SI), similar to the features observed among the scoriaceous and stubby particles. While dendritic microcrystalline textures are observed in terrestrial volcanic rocks that have undergone quench conditions, they are also identified as sudden quench features due to frictional heating for micrometeorites that have undergone atmospheric entry (Pandey et al., 2023), and may be worth closer inspection to verify their origins.

The cross-section of the scoriaceous particle 17NMAG-28 (**Figure 3e, d**), reveals the presence of a spinel rim. Sample 17NMAG-28 is suggested to be either extraterrestrial as spinel rims are characteristic of partially melted micrometeorites due to oxidation of ferrous iron in the silicate melt during interaction with atmospheric oxygen, and are observed in micrometeorites that consist of Fe-Ni-sulfides, metal beads, and silicates (Engrand et al., 2005; Feng et al., 2005; Folco and Cordier, 2015; Genge et al., 2008; Toppani et al., 2001; Toppani and Libourel, 2003; Van Maldeghem et al., 2023). While microtektites have also been characterized to exhibit magnetite rims, there are difficulties ascribing this particle to be an impact spherule, particularly one that belongs to the Australasian strewn field, owing to its lack of overlap in terms of bulk composition in the Mg-Si-Fe ternary diagram (**Figure 4a**). Other quench features such as dendritic structures consisting of Fe-Al spinels, are present throughout the cross-section. The Fe-Al oxide relict grains exhibit zoning (**Figure 3f**), where the dark regions are more abundant in Al and less abundant in Fe than those of the light regions (Loeb et al., 2024). Like other D-type spherules, the particle is suggested to be highly differentiated because of its low Mg-content (Loeb et al., 2024). The high-Fe content of this sample, in contrast to the low amount of Mg observed, suggests that if indeed the origin of this particle is extraterrestrial, this type of particle may be a compositionally unique occurrence among micrometeorites discussed thus far. Within the four morphological categories of the particles in this study, the scoriaceous D-type group is suggested to consist of particles that may be of extraterrestrial origin. However, further verification through other proxies such as oxygen isotope measurements may be necessary to verify their origins.

The stubby particle, IS19M-12 (**Figure 2j; Figure 3d, e, f**), is characterized by an aggregate of crystals composed of Fe-Al-oxides which are suggested to have been arranged circularly or spherically. Dendritic microcrystalline structures indicate quench conditions, similarly to what is observed in airburst particles (Van Ginneken et al., 2024) or in lavas. Similarly to the scoriaceous particle 17NMAG-28, there are problems directly related to identifying this particle to be a microtektite from the Australasian strewn field, due to the lack of overlap in major element composition with such identified materials (**Figure 4b**). Here, while the lack of a morphological feature such as a clear magnetite or spinel rim makes it difficult to establish an extraterrestrial origin, their compositional similarities to the scoriaceous particles may suggest a similar history.

The presence of magmaphile elements (e.g., Iyer et al., 1997a) and fractionated REE patterns (e.g., Folco and Cordier, 2015) have been used as discerning factors to distinguish spherules of terrestrial origin as opposed to cosmic spherules. However, these criteria are in general invalid, as they would rule out micrometeorites coming from the Moon or Mars. Assigning a terrestrial origin to a particle due to indications of a high degree of differentiation compared to chondrites is seemingly at odds with indications of atmospheric entry from the texture. Despite the presence of martian and lunar meteorites on the Earth, and the discovery of cosmic spherules from Vesta (Cordier et al., 2012, 2011) micrometeorites from larger planetary bodies such as the Moon and Mars have yet to be determined. Iron-rich and high-Ti lunar basalts suggest that D-type particles, which are characterized by high amounts of Fe, may be a candidate for the investigation of lunar and martian origins.

*D-type spherules and their correlations between morphologies and compositional subclasses*

Many of the D-type particles identified in Loeb et al. (2024) are highly depleted in Mg and are enriched in Fe and Si. While there are some similarities in major element composition between the specimens suggested to be of a terrestrial versus extraterrestrial origin, they are distinguished by their subtle morphological features. The blocky and vesicular D-type particles are determined to be objects of terrestrial origin. Meanwhile, the subcategories of the D-type particles characterized to be "scoriaceous" and "stubby" exhibit textures which are characteristic of quenching. In particular, one specimen from the scoriaceous subcategory exhibits a spinel rim. Dendritic microcrystalline structures composed of Fe-Al-oxides are observed for both stubby and scoriaceous particles. While these particles share commonalities with features that tend to be observed in micrometeorites, further investigation is desirable before their origins are decisively determined.

The morphologies of the particles are demonstrated to correspond to the various subclasses of D-type compositions to varying degrees. Many of the blocky particles in this collection consist of particles from the high-Sr, high-Si subgroup (**Figures 4c, 5**). In contrast, the vesicular particles mostly consist of the two subclassifications of low-Sr D-type subgroups (**Figures 4d, 5**). Meanwhile, there are specific subclasses of D-type particles that are exclusively observed among different morphological groupings. For instance, the high-Sr, low-Sr group is observed only among the scoriaceous and stubby particles, (**Figures 4b, 5**) whereas the low-Sr, high-Sr group, among the blocky and vesicular particles (**Figures 4c, d, 5**). While the low-Sr, low-Si subgroup is dispersed among the scoriaceous/stubby vs the vesicular particles (**Figures 4b, c, d, 5**), the Si-content observed among the vesicular particles are generally higher than the ones observed among the scoriaceous and stubby groups.

The correlation that is roughly observed between morphologies and composition may reflect the varying degrees of exposure to different temperature conditions. Due to the differences in volatility for Fe and Si, the Fe/Si ratios are observed to vary among cosmic spherules of varying morphologies that reflect the temperature conditions they were exposed to (Folco and Cordier, 2015; Taylor and Brownlee, 1991). The Fe/Si ratios decrease from PO spherules that are exposed to low peak temperatures compared to V-type spherules, which experienced higher peak temperatures. This suggests that the higher the exposure to high temperatures, the more the Fe/Si ratios will decrease in D-type particles, where the loss of Fe is favored during atmospheric entry. This further suggests that the Fe-enrichment of D-type particles that exhibit scoriaceous or stubby morphologies may not be due to shift of Fe/Si ratios due to exposure to high temperatures, but a reflection of the Fe-enrichment of their source compositions, where Fe/Si ratios would have subsequently shifted due to various thermal conditions.

# Summary and conclusion

The morphologies of a fraction of the D-type spherules in the CNEOS collection were studied through backscattered electron imaging and spot analysis of compositions and compared to the micro-XRF bulk compositional data from Loeb et al. (2024). The D-type particles of this collection suggest that the morphologies and sub-classes of D-type particles correlate with each other to a degree, where:

- The blocky particles of the D-type particles mostly consist of the high-Sr, high-Si subclass of D-type particles and are suggested to be of terrestrial origin.
- Vesicular particles largely consist of low-Sr, low-Si D-type spherules, and are suggested to be fragments of terrestrial basaltic lavas.
- The scoriaceous and stubby particles consist of an array of D-type particles mostly from the high-Sr, low-Si and low-Sr, low-Si groups.
- The lack of overlap of the bulk compositions to microtektites from the Australasian strewn field suggests it is difficult to attribute these particles to impact particles.
- The high-Sr, low-Si particles so far are only observed among the scoriaceous and stubby particles.
- The scoriaceous and stubby particles exhibit morphological textures suggestive of quenching, such as a spinel rim observed in one sample and dendritic microcrystalline structures. Among the four subcategories of D-type particles arranged according to their morphologies, these particles are suitable candidates for exploration of extraterrestrial origins.

The morphologies and compositions of the D-type particles are reflective of their origins and recent thermal history. The D-type particles in this study indicate that to a first approximation, the combination of morphologies and bulk composition demonstrate a degree of consistency that may be used as indicators for their source. The unusually high-Fe content of the D-type particles, along with their highly differentiated nature compared to achondrites suggest that these particles may be candidates for planetary precursors of rocky planets such as the Moon and Mars. However, more investigation is necessary before establishing their origins.

*Acknowledgements* We thank C. Hoskinson for funding the expedition, the Galileo Project at Harvard University for administrative and research support, and the Origins of Life Initiative at Harvard for supporting part of the analytical work on spherules. We additionally thank Michail I. Petaev for the EPMA imaging and analysis that were conducted in this study.

## References

Amonkar, A., Iyer, S.D., Babu, E.V.S.S.K., Shailajha, N., Sardar, A., Manju, S., 2021. Fluid-driven hydrovolcanic activity along fracture zones and near seamounts: Evidence from deep-sea Fe-rich spherules, Central Indian Ocean Basin. Acta Geol. Sin. (English Ed. 95, 1591–1603. https://doi.org/10.1111/1755-6724.14697

Anders, E., Grevesse, N., 1989. Abundances of the elements: Meteoritic and solar. Geochim. Cosmochim. Acta 53, 197–214. https://doi.org/10.1016/0016-7037(89)90286-X

Blanchard, M.B., Brownlee, D.E., Bunch, T.E., Hodge, P.W., Kyte, F.T., 1980. Meteoroid ablation spheres from deep-sea sediments. Earth Planet. Sci. Lett. 46, 178–190. https://doi.org/10.1016/0012-821X(80)90004-7

Brase, L.E., Harvey, R., Folco, L., Suttle, M.D., McIntosh, E.C., Day, J.M.D., Corrigan, C.M., 2021. Microtektites and glassy cosmic spherules from new sites in the Transantarctic

Mountains, Antarctica. Meteorit. Planet. Sci. 56, 829–843. https://doi.org/10.1111/maps.13634

Carracedo-Sánchez, M., Sarrionandia, F., Arostegui, J., Errandonea-Martin, J., Gil-Ibarguchi, J.I., 2016. Petrography and geochemistry of achnelithic tephra from Las Herrerías Volcano (Calatrava volcanic field, Spain): Formation of nephelinitic achneliths and post-depositional glass alteration. J. Volcanol. Geotherm. Res. 327, 484–502. https://doi.org/10.1016/j.jvolgeores.2016.09.006

Cordier, C., Folco, L., Taylor, S., 2011. Vestoid cosmic spherules from the South Pole Water Well and Transantarctic Mountains (Antarctica): A major and trace element study. Geochim. Cosmochim. Acta 75, 1199–1215. https://doi.org/10.1016/j.gca.2010.11.024

Cordier, C., Suavet, C., Folco, L., Rochette, P., Sonzogni, C., 2012. HED-like cosmic spherules from the Transantarctic Mountains, Antarctica: Major and trace element abundances and oxygen isotopic compositions. Geochim. Cosmochim. Acta 77, 515–529. https://doi.org/10.1016/j.gca.2011.10.021

Engrand, C., McKeegan, K.D., Leshin, L.A., Herzog, G.F., Schnabel, C., Nyquist, L.E., Brownlee, D.E., 2005. Isotopic compositions of oxygen, iron, chromium, and nickel in cosmic spherules: Toward a better comprehension of atmospheric entry heating effects. Geochim. Cosmochim. Acta 69, 5365–5385. https://doi.org/10.1016/j.gca.2005.07.002

Feng, H., Jones, K.W., Tomov, S., Stewart, B., Herzog, G.F., Schnabel, C., Brownlee, D.E., 2005. Internal structure of type I deep-sea spherules by X-ray computed microtomography. Meteorit. Planet. Sci. 40, 195–206. https://doi.org/10.1111/j.1945-5100.2005.tb00375.x

Folco, L., Cordier, C., 2015. Micrometeorites. Eur. Mineral. Union Notes Mineral. 15, 253–297. https://doi.org/10.1180/EMU-notes.15.9

Folco, L., D'Orazio, M., Tiepolo, M., Tonarini, S., Ottolini, L., Perchiazzi, N., Rochette, P., Glass, B.P., 2009. Transantarctic Mountain microtektites: Geochemical affinity with Australasian microtektites. Geochim. Cosmochim. Acta 73, 3694–3722. https://doi.org/10.1016/j.gca.2009.03.021

Folco, L., Glass, B.P., D'Orazio, M., Rochette, P., 2018. Australasian microtektites: Impactor identification using Cr, Co and Ni ratios. Geochim. Cosmochim. Acta 222, 550–568. https://doi.org/10.1016/j.gca.2017.11.017

Folco, L., Perchiazzi, N., D'Orazio, M., Frezzotti, M.L., Glass, B.P., Rochette, P., 2010. Shocked quartz and other mineral inclusions in Australasian microtektites. Geology 38, 211–214. https://doi.org/10.1130/G30512.1

Folco, L., Rochette, P., D'Orazio, M., Masotta, M., 2023. The chondritic impactor origin of the Ni-rich component in Australasian tektites and microtektites. Geochim. Cosmochim. Acta 360, 231–240. https://doi.org/10.1016/j.gca.2023.09.018

Genareau, Kimberly, Hong, Y.K., Lee, W., Choi, M., Rostaghi-Chalaki, M., Gharghabi, P., Gafford, J., Klüss, J., 2019. Effects of lightning on the magnetic properties of volcanic ash. Sci. Rep. 9, 4726. https://doi.org/10.1038/s41598-019-41265-3

Genareau, K., Wallace, K.L., Gharghabi, P., Gafford, J., 2019. Lightning effects on the grain size distribution of volcanic ash. Geophys. Res. Lett. 46, 3133–3141. https://doi.org/10.1029/2018GL081298

Genareau, K., Wardman, J.B., Wilson, T.M., McNutt, S.R., Izbekov, P., 2015. Lightning-induced volcanic spherules. Geology 43, 319–322. https://doi.org/10.1130/G36255.1

Genge, M.J., Engrand, C., Gounelle, M., Taylor, S., 2008. The classification of micrometeorites. Meteorit. Planet. Sci. 43, 497–515. https://doi.org/10.1111/j.1945-5100.2008.tb00668.x

Glass, B.P., Burns, C., 1988. Microkrystites-A new term for impact-produced glassy spherules containing primary crystallites. Lunar Planet. Sci. Conf. 18th 455–458.

Glass, B.P., Huber, H., Koeberl, C., 2004. Geochemistry of Cenozoic microtektites and clinopyroxene-bearing spherules. Geochim. Cosmochim. Acta 68, 3971–4006. https://doi.org/10.1016/j.gca.2004.02.026

Glass, B.P., Simonson, B.M., 2012. Distal impact ejecta layers: Spherules and more. Elements 8, 43–48. https://doi.org/10.2113/gselements.8.1.43

Goresy, A. El, 1968. Electron microprobe analysis and ore microscopic study of magnetic spherules and grains collected from the Greenland ice. Contrib. to Mineral. Petrol. 17, 331–346. https://doi.org/10.1007/BF00380743

Iwahashi, J., 1991. Shape and Surface Structure of the Magnetic Micro-Spherules from Permian and. J. Geosci. 34, 55–73.

Iyer, S.D., Prasad, M.S., Gupta, S.M., Charan, S.N., 1997a. Evidence for recent hydrothermal activity in the Central Indian Basin. Deep. Res. Part I Oceanogr. Res. Pap. 44, 1167–1173. https://doi.org/10.1016/S0967-0637(97)00001-0

Iyer, S.D., Prasad, M.S., Gupta, S.M., Charan, S.N., Mukherjee, A.D., 1997b. Hydrovolcanic activity in the Central Indian Ocean Basin. Does nature mimic laboratory experiments? J. Volcanol. Geotherm. Res. 78, 209–220. https://doi.org/10.1016/S0377-0273(97)00017-6

Loeb, A., Jacobsen, S.B., Tagle, R., Adamson, T., Bergstrom, S., Cherston, J., Peddeti, C., Pugh, J., Samuha, S., Sasselov, D.D., Schlereth, M., Siler, J., Siraj, A., 2024. Chemical classification of spherules recovered from the Pacific Ocean site of the CNEOS 2014-01-08 (IM1) bolide. Chem. Geol. 670, 122415. https://doi.org/10.1016/j.chemgeo.2024.122415

McDonough, W.F., Sun, S. -s., 1995. The composition of the Earth. Chem. Geol. 120, 223–253. https://doi.org/10.1016/0009-2541(94)00140-4

Moune, S., Faure, F., Gauthier, P.J., Sims, K.W.W., 2007. Pele's hairs and tears: Natural probe of volcanic plume. J. Volcanol. Geotherm. Res. 164, 244–253. https://doi.org/10.1016/j.jvolgeores.2007.05.007

Noguchi, T., Yabuta, H., Itoh, S., Sakamoto, N., Mitsunari, T., Okubo, A., Okazaki, R., Nakamura, T., Tachibana, S., Terada, K., Ebihara, M., Imae, N., Kimura, M., Nagahara, H., 2017. Variation of mineralogy and organic material during the early stages of aqueous activity recorded in Antarctic micrometeorites. Geochim. Cosmochim. Acta 208, 119–144. https://doi.org/10.1016/j.gca.2017.03.034

Pandey, M., Rudraswami, N.G., Singh, V.P., Viegas, A., 2023. Geochemical evaluation of cosmic spherules collected from the Central Indian Ocean Basin. Deep. Res. Part I Oceanogr. Res. Pap. 200, 104153. https://doi.org/10.1016/j.dsr.2023.104153

Petaev, M.I., Jacobsen, S.B., 2009. Petrologic study of SJ101, a new forsterite-bearing CAI from the Allende CV3 chondrite. Geochim. Cosmochim. Acta 73, 5100–5114. https://doi.org/10.1016/j.gca.2008.10.045

Porritt, L.A., Russell, J.K., Quane, S.L., 2012. Pele's tears and spheres: Examples from Kilauea Iki. Earth Planet. Sci. Lett. 333–334, 171–180. https://doi.org/10.1016/j.epsl.2012.03.031

Rudnick, R.L., Gao, S., 2003. Composition of the Continental Crust, 2nd ed, Treatise on Geochemistry. https://doi.org/10.1016/B978-0-08-095975-7.00301-6

Simonson, B.M., Glass, B.P., 2004. Spherule layers - records of ancient impacts. Annu. Rev. Earth Planet. Sci. 32, 329–361. https://doi.org/10.1146/annurev.earth.32.101802.120458

Soens, B., Chernonozhkin, S.M., González de Vega, C., Vanhaecke, F., van Ginneken, M., Claeys, P., Goderis, S., 2022. Characterization of achondritic cosmic spherules from the Widerøefjellet micrometeorite collection (Sør Rondane Mountains, East Antarctica). Geochim. Cosmochim. Acta 325, 106–128. https://doi.org/10.1016/j.gca.2022.03.029

Taylor, S., Brownlee, D.E., 1991. Cosmic spherules in the geologic record. Meteorit. Planet. Sci. 26, 203–211.

Taylor, S., Herzog, G.F., Delaney, J.S., 2007. Crumbs from the crust of Vesta: Achondritic cosmic spherules from the South Pole water well. Meteorit. Planet. Sci. 42, 223–233. https://doi.org/10.1111/j.1945-5100.2007.tb00229.x

Taylor, S., Lever, J.H., Harvey, R.P., 2000. Numbers, types, and compositions of an unbiased collection of cosmic spherules. Meteorit. Planet. Sci. 35, 651–666. https://doi.org/10.1111/j.1945-5100.2000.tb01450.x

Toppani, A., Libourel, G., 2003. Factors controlling compositions of cosmic spinels: Application to atmospheric entry conditions of meteoritic materials. Geochim. Cosmochim. Acta 67, 4621–4638. https://doi.org/10.1016/S0016-7037(03)00383-1

Toppani, A., Libourel, G., Engrand, C., Maurette, M., 2001. Experimental simulation of atmospheric entry of micrometeorites. Meteorit. Planet. Sci. 36, 1377–1396. https://doi.org/10.1111/j.1945-5100.2001.tb01831.x

Van Ginneken, M., Gattacceca, J., Rochette, P., Sonzogni, C., Alexandre, A., Vidal, V., Genge, M.J., 2017. The parent body controls on cosmic spherule texture: Evidence from the oxygen isotopic compositions of large micrometeorites. Geochim. Cosmochim. Acta 212, 196–210. https://doi.org/10.1016/j.gca.2017.05.008

Van Ginneken, M., Genge, M.J., Harvey, R.P., 2018. A new type of highly-vaporized microtektite from the Transantarctic Mountains. Geochim. Cosmochim. Acta 228, 81–94. https://doi.org/10.1016/j.gca.2018.02.041

Van Ginneken, M., Harvey, R.P., Goderis, S., Artemieva, N., Boslough, M., Maeda, R., Gattacceca, J., Folco, L., Yamaguchi, A., Sonzogni, C., Wozniakiewicz, P., 2024. The identification of airbursts in the past: Insights from the BIT-58 layer. Earth Planet. Sci. Lett. 627. https://doi.org/10.1016/j.epsl.2023.118562

Van Ginneken, M. Van, Penelope, J., Brownlee, D.E., Corte, V. Della, Delauche, L., Duprat, J., Engrand, C., Folco, L., Fries, M., Gattacceca, J., Matthew, J., Goderis, S., Gounelle, M., Harvey, R.P., Jonker, G., Krämer, L., Larsen, J., Lever, J.H., Rojas, J., Rotundi, A., Maldeghem, F. Van, 2024. Micrometeorite collections: a review and their current status. Philos. Trans. R. Soc. A 382, 20230195.

Van Maldeghem, F., van Ginneken, M., Soens, B., Kaufmann, F., Lampe, S., Krämer Ruggiu, L., Hecht, L., Claeys, P., Goderis, S., 2023. Geochemical characterization of scoriaceous and unmelted micrometeorites from the Sør Rondane Mountains, East Antarctica: Links to chondritic parent bodies and the effects of alteration. Geochim. Cosmochim. Acta 354, 88–108. https://doi.org/10.1016/j.gca.2023.06.002

Wozniakiewicz, P.J., Alesbrook, L.S., Bradley, J.P., Ishii, H.A., Price, M.C., Zolensky, M.E., Brownlee, D.E., van Ginneken, M., Genge, M.J., 2024. Atmospheric collection of extraterrestrial dust at the Earth's surface in the mid-Pacific. Meteorit. Planet. Sci. 2817, 2789–2817. https://doi.org/10.1111/maps.14251

Yan, P., Xiao, Z., Xiao, G., Pan, Q., Hui, H., Wu, Y., Ma, Y., Xu, Y., 2022. Undetection of Australasian microtektites in the Chinese Loess Plateau. Palaeogeogr. Palaeoclimatol. Palaeoecol. 585, 110721. https://doi.org/10.1016/j.palaeo.2021.110721